\documentclass[traditabstract]{aa}
\usepackage{graphicx}
\usepackage{natbib}
\usepackage{txfonts}
\usepackage{ulem}
      \def\new#1 {{\bf #1 }}
      \def\cut#1 {\sout{#1} }

\def\cmsq  {$\hbox{{\rm cm}}^{-2}$}    

\def\CHP {CH$^+$}
\def\THCHP {$^{13}$CH$^+$}
\def\OHP {OH$^+$}
\def\SHP {SH$^+$}
\def\HTOP {H$_3$O$^+$}

\def\OI{O{\sc i}}

\def\folio{\ifnum\pageno=1\nopagenumbers\else\number\pageno\fi}

\def\lax    {\ifmmode{_<\atop^{\sim}}\else{${_<\atop^{\sim}}$}\fi}
\def\gax    {\ifmmode{_>\atop^{\sim}}\else{${_>\atop^{\sim}}$}\fi}
\newbox\grsign      \setbox\grsign=\hbox{$>$} 
\newdimen\grdimen   \grdimen=\ht\grsign
\newbox\simgreatbox \setbox\simgreatbox=\hbox{\raise.5ex\hbox{$>$}\llap
                        {\lower.5ex\hbox{$\sim$}}}\ht1=\grdimen\dp1=0pt
\newbox\simlessbox  \setbox\simlessbox =\hbox{\raise.5ex\hbox{$<$}\llap
                        {\lower.5ex\hbox{$\sim$}}}\ht2=\grdimen\dp2=0pt

\newbox\grsign \setbox\grsign=\hbox{$>$} \newdimen\grdimen \grdimen=\ht\grsign
\newbox\laxbox \newbox\gaxbox
\setbox\gaxbox=\hbox{\raise.5ex\hbox{$>$}\llap
     {\lower.5ex\hbox{$\sim$}}}\ht1=\grdimen\dp1=0pt
\setbox\laxbox=\hbox{\raise.5ex\hbox{$<$}\llap
     {\lower.5ex\hbox{$\sim$}}}\ht2=\grdimen\dp2=0pt
\def\gax{\mathrel{\copy\gaxbox}}
\def\lax{\mathrel{\copy\laxbox}}
\def\boxit#1    {\vbox{\hrule\hbox{\vrule\kern3pt
                  \vbox{\kern3pt#1\kern3pt}\kern3pt\vrule}\hrule}}
\def\h      {\ifmmode{^{\rm h}}\else{$^{\rm h}$}\fi}
\def\m      {\ifmmode{^{\rm m}}\else{$^{\rm m}$}\fi}
\def\s      {\ifmmode{^{\rm s}}\else{$^{\rm s}$}\fi}
\def\decas    {\ifmmode{{\rlap.}{''}}\else{${\rlap.}{''}$}\fi}
\def\mum     {\ifmmode{\mu{\rm m}}\else{$\mu{\rm m}$}\fi}
\def\s      {\ifmmode{^{\rm s}}\else{$^{\rm s}$}\fi}
\def\deg      {\ifmmode{^{\circ}}\else{$^{\circ}$}\fi}
\def\as     {\ifmmode {\rlap.}$\,$''$\,$\! \else ${\rlap.}$\,$''$\,$\!$\fi}
\def\decsec  {\ifmmode {\rlap.}$\,$^{s}$\,$\! \else ${\rlap.}$\,$^{s}$\,$\!$\fi}\def\decs  {\ifmmode {\rlap.}$\,$^{s}$\,$\! \else ${\rlap.}$\,$^{s}$\,$\!$\fi}

\def\kms    {\ifmmode{{\rm km~s}^{-1}}\else{km~s$^{-1}$}\fi}

\def\Mspy   {\ifmmode {M_{\odot} {\rm yr}^{-1}} \else $M_{\odot}$~yr$^{-1}$\fi}
\def\Mdot   {\ifmmode {\dot M} \else $\dot M$\fi}
\def\mhd    {\ifmmode {n_{{\rm H}_2}} \else $n_{{\rm H}_2}$\fi}
\def\mhcd   {\ifmmode {N_{{\rm H}_2}} \else $N_{{\rm H}_2}$\fi}

\def\El      {\ifmmode{E_{\ell}}\else{$E_{\ell}$}\fi}
\def\beam    {\ifmmode{\theta_{\rm B}}\else{$\theta_{\rm B}$}\fi}
\def\mjyb   {\ifmmode {{\rm mJy~beam}^{-1}} \else{mJy~beam$^{-1}$}\fi}
\def\mujyb   {\ifmmode {\mu{\rm Jy~beam}^{-1}} \else{$\mu$Jy~beam$^{-1}$}\fi}
\def\Trot   {\ifmmode{T_{\rm rot}}\else$T_{\rm rot}$\fi}    
    
\def\Teff   {\ifmmode{T_{\rm eff}}\else$T_{\rm eff}$\fi}

\def\ITRS   {\ifmmode{\smallint {\rm T}_{R}^{*}dv}\else{$\smallint 
{\rm T}_{R}^{*}dv$}\fi}
\def\ITRS   {\ifmmode{\smallint {\rm T}_{R}^{*}dv}\else{$\smallint 
{\rm T}_{R}^{*}dv$}\fi}
\def\ITAS   {\ifmmode{\smallint {\rm T}_{A}^{*}dv}\else{$\smallint 
{\rm T}_{A}^{*}dv$}\fi}

\def\OI         {O~{\eightpt I}}

\def\lefttitle#1  {\noindent \hangindent=18.0pt \hangafter=1 {#1} \par}
\def\vol#1  {{\bf {#1}{\rm,}\ }}
\font\eightpt=cmr8

\font\tenssb=cmssbx10
\font\tenbf=cmbx10
\font\sevenbf=cmbx8
\font\fivebf=cmbx6
\def\unetdemi    {\smallskipamount=6pt plus2pt minus2pt
                  \medskipamount=12pt plus4pt minus4pt
                  \bigskipamount=24pt plus8pt minus8pt
                  \normalbaselineskip=16pt plus0pt minus0pt
                  \normallineskip=2pt
                  \normallineskiplimit=0pt
                  \jot=6pt
                  {\def\smallskip {\vskip\smallskipamount}}
                  {\def\medskip   {\vskip\medskipamount}}
                  {\def\bigskip   {\vskip\bigskipamount}}
                  {\setbox\strutbox=\hbox{\vrule 
                    height17.0pt depth7.0pt width 0pt}}
                  \parskip 12.0pt
                  \normalbaselines}
\def\smallerspace {\smallskipamount=3pt plus0pt minus0pt
                  \medskipamount=6pt plus0pt minus0pt
                  \bigskipamount=10.5pt plus0pt minus0pt
                  \normalbaselineskip=10.5pt plus0pt minus0pt
                  \normallineskip=1pt
                  \normallineskiplimit=0pt
                  \jot=3pt
                  {\def\smallskip {\vskip\smallskipamount}}
                  {\def\medskip   {\vskip\medskipamount}}
                  {\def\bigskip   {\vskip\bigskipamount}}
                  {\setbox\strutbox=\hbox{\vrule 
                    height8.5pt depth3.5pt width 0pt}}
                  \parskip 0pt
                  \normalbaselines}
\def\memospace    {\smallskipamount=4pt plus1pt minus1pt
                  \medskipamount=6pt plus2pt minus2pt
                  \bigskipamount=14pt plus6pt minus6pt
                  \normalbaselineskip=14pt plus0pt minus0pt
                  \normallineskip=1pt
                  \normallineskiplimit=0pt
                  \jot=4pt
                  {\def\smallskip {\vskip\smallskipamount}}
                  {\def\medskip   {\vskip\medskipamount}}
                  {\def\bigskip   {\vskip\bigskipamount}}
                  {\setbox\strutbox=\hbox{\vrule 
                    height17.0pt depth7.0pt width 0pt}}
                  \parskip 2.0pt
                  \normalbaselines}
\def\memowidespace    {\smallskipamount=5pt plus1pt minus1pt
                  \medskipamount=7.5pt plus2pt minus2pt
                  \bigskipamount=17.5pt plus6pt minus6pt
                  \normalbaselineskip=17.0pt plus0pt minus0pt
                  \normallineskip=1.25pt
                  \normallineskiplimit=0pt
                  \jot=5pt
                  {\def\smallskip {\vskip\smallskipamount}}
                  {\def\medskip   {\vskip\medskipamount}}
                  {\def\bigskip   {\vskip\bigskipamount}}
                  {\setbox\strutbox=\hbox{\vrule 
                    height21.25pt depth8.75pt width 0pt}}
                  \parskip 2.5pt
                  \normalbaselines}
\begin{document}

\title{First interstellar detection of \OHP}
\author{F. Wyrowski \inst{1} \and 
        K. M. Menten \inst{1} \and 
        R. G\"usten \inst{1} \and 
        A. Belloche \inst{1}  }

\offprints{F. Wyrowski}

\institute{Max-Planck-Institut f\"ur Radioastronomie,
Auf dem H\"ugel 69, D-53121 Bonn, Germany\\
\email{wyrowski, kmenten, rguesten, belloche@mpifr-bonn.mpg.de}
}

\date{Received / Accepted}

\titlerunning{Interstellar \OHP}

\authorrunning{Wyrowski et al.}

\abstract {
  The Atacama Pathfinder Experiment (APEX) 12 m telescope was used to
  observe the $N=1-0, J=0-1$ ground-state transitions of \OHP\ at
  909.1588~GHz with the CHAMP+ heterodyne array receiver.  Two blended
  hyperfine structure transitions were detected in absorption against
  the strong continuum source Sagittarius B2(M) and in several pixels
  offset by 18\arcsec.  Both absorption from Galactic center gas and
  absorption from diffuse clouds in intervening spiral arms in
  a velocity range from --116 to 38.5~\kms\ is observed.  The
    total \OHP\ column density of absorbing gas is $2.4\times
    10^{15}$~\cmsq. A column
  density local to Sgr B2(M) of $2.6 \times 10^{14}$~\cmsq\ is
  found. On the intervening line-of-sight, the column density per unit
  velocity interval is in the range of 1 to $40 \times
  10^{12}$~\cmsq/(\kms). \OHP\ is found to be on average more abundant
  than other hydrides, such as \SHP\ and \CHP. Abundance ratios of OH
  and atomic oxygen to \OHP\ are found in the range of
  $10^{1-2}$ and $10^{3-4}$, respectively. The detected absorption of
  a continuous velocity range on the line-of-sight shows \OHP\ to be
  an abundant component of diffuse clouds.
}


\maketitle

\section{\label{intro}Introduction}

Hydrides are key ingredients of interstellar chemistry since they are
the initial products of chemical networks that lead to the formation
of more complex molecules. The fundamental rotational transitions of
light hydrides fall into the submillimeter wavelength range. A so far
elusive but nevertheless important hydride is \OHP,
oxoniumylidene. \OHP\ has a $^3\Sigma^-$ ground electric state. Its
$N=1-0$ transition near 1~THz is split into three fine structure
components, which are further split into several hyperfine components
\citep{bekooy+1985,mueller+2005}.

First estimates of \OHP\ abundances were given in the calculations of
\citet{glassgold_langer1976} and \citet{barsuhn_walmsley1977}. The
latter study predicts the highest abundances in low density gas.
\citet{dealmeida+1981} conclude that the \OHP\ molecule is primarily
formed in warm, diffuse and moderately thick interstellar clouds. 
  Later \citet{dealmeida1990} discusses the relevance of shocks for
  the production of \OHP\ and gives a comprehensive summary of the
  fine and hyperfine rotational structure and transition probabilities
  of \OHP. 

\citet{polehampton+2007} searched for the $N=2-1, J=1-2$ and $
  N=2-1$, $J=3-2$ transitions at 153.47 and 152.9897~\mum,
respectively, in the ISO LWS spectrum of Sgr B2 but found a spectral
feature at the wavelength of only one of the lines and therefore ruled
out a detection of \OHP. Also \citet{gonzalez+2004} discuss an
absorption feature in the ISO LWS spectrum of Arp 220 at 153~\mum\ but
conclude that it belongs to NH.

A detection of \OHP\ with the ion mass spectrometer aboard the Giotto
spacecraft in the tail of the comet was reported by
\citet{balsiger1986}, where \OHP\ was found to be one of the main
cometary hot-ion species in the coma.

\OHP\ plays an important role also in the formation of water. 
  \OHP\ is rapidly converted into \HTOP\ which then leads to the
  formation of water and OH \citep[see e.g.][]{wannier+1991}.  This
holds in cold and warm environments likewise. Hence, it can either be
created via $\rm{H}_3^+$ by

\begin{displaymath}
  {\rm H_3^+ + O \rightarrow  H_2 + OH}^+
\end{displaymath}
or in warmer regions via H$^+$ with the following two reactions:
\begin{displaymath}
  {\rm H^+ + O \rightarrow  H + O^+} - k\times 232~K,
\end{displaymath}
\begin{displaymath}
  {\rm O^+ + H_2 \rightarrow  OH^+ + H}.
\end{displaymath}
The \OHP\ then reacts rapidly with molecular hydrogen:
\begin{displaymath}
  {\rm OH^+ + H_2 \rightarrow  H_2O^+ + H }.
\end{displaymath}
${\rm H_2O^+}$ subsequently reacts with molecular hydrogen to give
${\rm H_3O^+}$ which recombines then to form both, OH and water.
Therefore measurements of \OHP\ can provide crucial insight into the
part of the oxygen chemical network that leads to the formation of
water, which is currently a prominent target for observations with the
Herschel Space Observatory. In addition to \OHP, several other
hydrides (\SHP, \THCHP, and HCl) have recently been studied from the
ground in the accompanying paper by \citet{menten+2010} in the
absorbing diffuse clouds towards Sgr B2(M). This prompted us to also
search for \OHP and here, we describe the first detection of the \OHP\
ion in interstellar space using the Atacama Pathfinder Experiment
telescope (APEX).
\begin{figure*}[ht]
\begin{center}
\includegraphics[width=7.5cm,angle=-90]{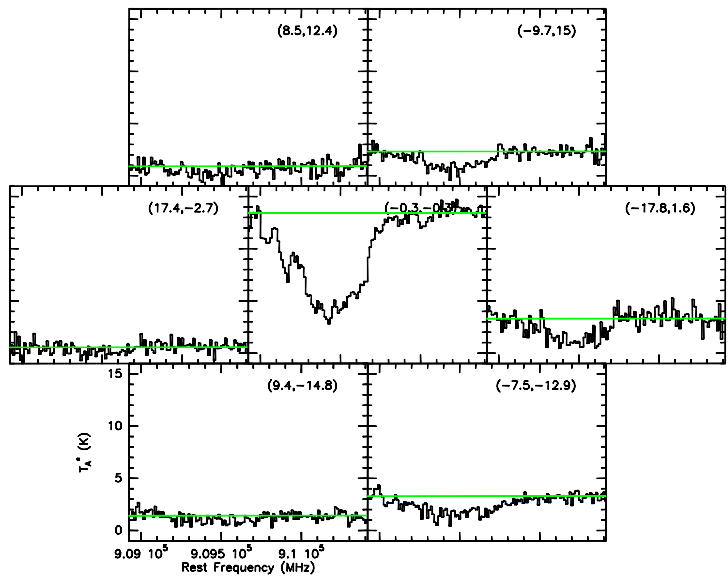}
\caption{\label{fig:hex7} \OHP\ spectrum towards Sgr B2(M) observed
  with the 7 pixels of the CHAMP+ high frequency array. The continuum
  level for each spectrum is indicated by a straight, green line and
  the offsets (Ra./Dec.) from Sgr B2(M) of the individual pixels are given in
  each box in second of arc. }
\end{center}
\end{figure*}

\section{\label{obs}Observations and data reduction}

The observations were carried out in August 2009 with the the 12-m
Atacama Pathfinder Experiment telescope (APEX\footnote{This
  publication is based on data acquired with the Atacama Pathfinder
  Experiment (APEX). APEX is a collaboration between the
  Max-Planck-Institut f\"ur Radioastronomie, the European Southern
  Observatory, and the Onsala Space Observatory.})
\citep{Gusten_etal2006} using the MPIfR-built CHAMP+ receiver array
\citep{Kasemann2006}. The stronger of the two hyperfine lines of the
\OHP\ $N=1-0, J=0-1$ transition at 909.1588~GHz was tuned into 
  the center of the upper sideband of the high frequency array of the
receiver. The array is operated in single sideband mode with a typical
rejection of 10db and consists out of seven pixels in a hexagonal
pattern with separations of about 18\arcsec\ and individual beam sizes
of 7\arcsec. The telescope was pointed towards Sgr B2(M) at a position
of $\alpha$(J2000)=$17^h47^m20.15^s$ and
$\delta$(J2000)=$ -28\deg\, 23\arcmin\, 04.9\arcsec $.  The SSB system
temperatures during the observations were in the range from 10000 to
20000~K for the individual pixels with a precipitable water vapor of
about 0.6~mm. The relatively high system temperatures are a result of
the nearby atmospheric water absorption at 916~GHz. As backends, MPIfR
Fast Fourier Transform spectrometers \citep{Klein_etal2006} were used
to cover a bandwidths of 2x1.5~GHz for each pixel with a velocity
resolution of 0.25~km/s that was later lowered to 2~km/s to increase
the signal-to-noise in the spectra. In the following, only spectra
from one FFTS per pixel are discussed to exclude possible platforming
effects.  The wobbling secondary was chopped with a frequency of
1.5~Hz and a throw of $120''$ about the cross elevation axis in a
symmetric mode which allows a reliable detection of the source
continuum as well. Pointing and focus corrections were determined with
cross scans on the source itself. With the online calibration, spectra
were calibrated to a $T_{\rm A}^*$ scale. The conversion to $T_{\rm
  MB}$ was done using a forward efficiency of 0.95 and a beam
efficiency measured on Mars at 809 GHz earlier in the month and then
scaled to 0.33 for the \OHP\ frequency.

\section{Results}
\label{sec:results}

\begin{figure}[ht]
\begin{center}
\includegraphics[height=0.48\textwidth,angle=-90]{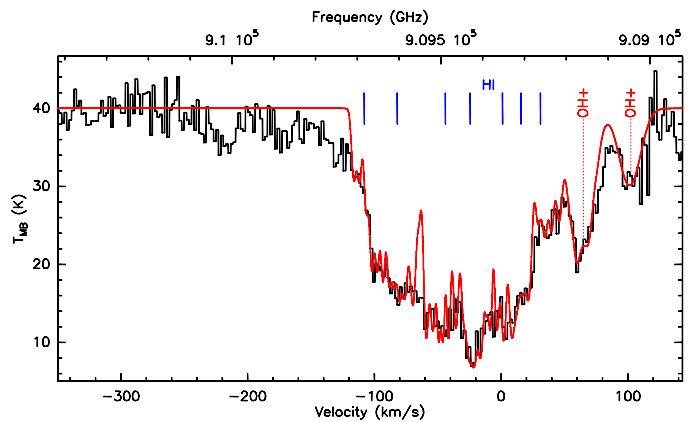}
\caption{\label{fig:spectrum} Observed \OHP\ spectrum towards Sgr
  B2(M) in black with the velocity scale referred to the stronger
  hyperfine line of \OHP. Frequencies for the two HFS lines at a
  velocity of 64~\kms\ local to Sgr B2(M) are indicated.  Also
  velocities of the HI components from \citet{garwood_dickey1989} are
  marked. The result from the multi-velocity component fit is overlaid
  in red.}
\end{center}
\end{figure}

\begin{table}
  \caption{Parameters of the observed lines of the $N=1-0$ transition. 
    The frequencies were taken from \citet{bekooy+1985} and the 
    Einstein $A$ coefficients from the Cologne Database for Molecular 
      Spectroscopy \citep[CDMS,][]{mueller+2005}.
      Note that the Einstein $A$ coefficients given by \citet{dealmeida1990}
      are too high by a factor 1.5.
  }
\label{t:parameters} 
\begin{center}
 \begin{tabular}{cccccc}
 \hline\hline 
$ J^{\prime} \leftarrow J^{\prime\prime}$ &
$ F^{\prime} \leftarrow F^{\prime\prime}$ & Frequency & $g_l$ & $g_u$ & A \\
  &   & (MHz) & & & ($10^{-2}s^{-1}$) \\ 
\hline 

0 1  & 1/2 1/2 & 909045.2 & 2 & 2  &  0.524 \\
0 1  & 1/2 3/2 & 909158.8 & 4 & 2  &  1.048 \\

\hline
 \end{tabular}
 \end{center}
\end{table}

Figure \ref{fig:hex7} shows the spectra towards Sgr B2(M) observed
with the 7 different pixels of the array. The baseline offset in the
spectra represents the continuum level of the source at the observed
positions and is by far strongest in the center pixel on the
source. But even towards offset pixels about 18\arcsec\ away from the
center of the source some continuum is detected. 
In the center
  pixel the continuum level is at 40~K, although with a slightly
  different choice of baseline, values as low as 36~K cannot be
  excluded (see Fig.~\ref{fig:spectrum}). 
The overall strength and
morphology of the continuum is comparable to that found in the
observations presented in the \SHP\ analysis of
\citet{menten+2010}. While the absorption from \OHP\ is most clearly
seen towards the center, it is also evident in some of the offset
pixels, with a smaller signal-to-noise though, showing that the \OHP\
absorption is spatially resolved.

The spectrum towards the center pixel is also shown in
Figure~\ref{fig:spectrum}. Here the frequencies of the two hyperfine
lines, which are given in Table~\ref{t:parameters}, are indicated
(referenced to a velocity of Sgr B2(M) of 64~\kms) and a velocity
scale is given with respect to the frequency of the stronger \OHP\
component. Velocities of atomic hydrogen velocity components that were
measured by \citet{garwood_dickey1989} on the line-of-sight towards
Sgr B2(M) are indicated for comparison.


The optical depth of the absorption can be estimated from the
  line-to-continuum ratio at each velocity as $\tau = -\log(1-T_{\rm
    L}/T_{\rm C})$.  This leads to typical opacities of 1, going up to
  1.5 at a velocity of --20~\kms. The total, velocity integrated
  optical depth of the \OHP\ absorption from both blended hyperfine
  lines is $\tau\Delta v=175$~\kms.

\section{Analysis}
\label{s:analysis}

\begin{figure}[t]
\begin{center}
\includegraphics[height=0.48\textwidth,angle=-90]{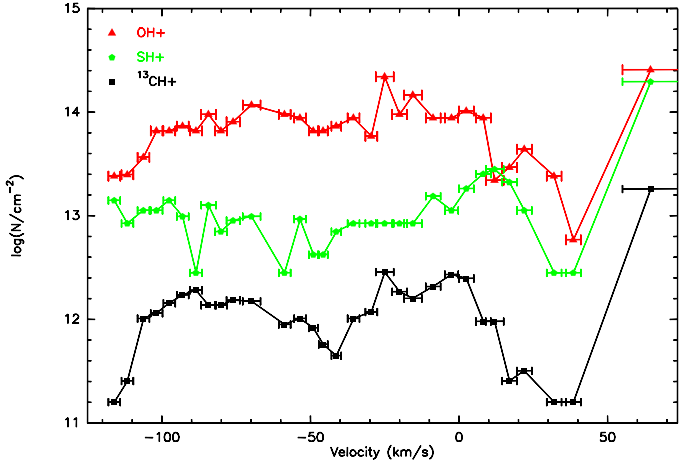}
\caption{\label{fig:cd-comparison} Comparison of hydride column
  densities from this work and \citet{menten+2010} as a function of
  $v_{\rm LSR}$. The velocity widths of the components are indicated by
  horizontal bars.}
\end{center}
\end{figure}

\begin{figure}[t]
\begin{center}
\includegraphics[height=0.48\textwidth,angle=-90]{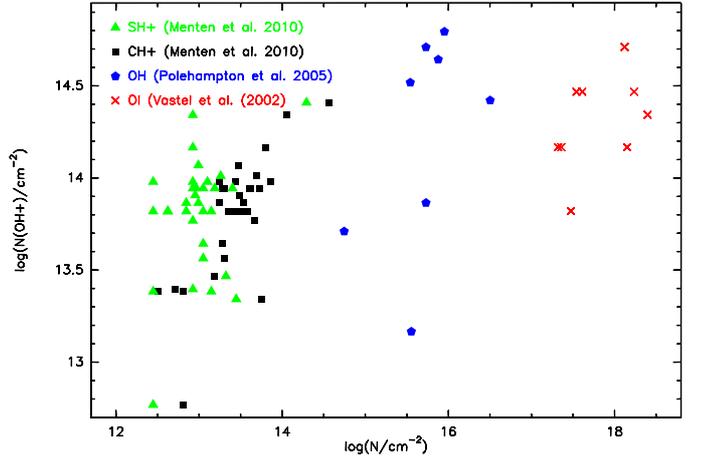}
\caption{\label{fig:ratios} Correlations of column densities from
  different molecules with \OHP. \CHP\ column densities were computed
  from \THCHP\ using the isotopic ratios given in
  \citet{menten+2010}. Note, since OH and \OI\ were taken from ISO
  observations with low velocity resolution, fewer and broader
  velocity components were used to determine the corresponding \OHP\
  column densities.}
\end{center}
\end{figure}

From the integrated optical depths the total column density
of the absorbing \OHP\ can be computed from \citep[e.g.][]{comito+2003}:

\begin{equation}
     N{_{\rm T}} = \frac{8\pi\nu^3}{A_{ul}c^3}\,\frac{g_l}{g_u}\,\frac{1}{f}\,\tau\Delta v
\end{equation} 

Here it is assumed that only the ground state of the molecule is
populated. The $A_{ul}$ are given in Table~\ref{t:parameters}
and are based on the electric dipole
moment that \citet{werner+1983} obtained from ab initio calculation.
$f$ is the fraction of molecules in the corresponding ground state and $g_l$
and $g_u$ are the statistical weights of the lower and upper levels, respectively. For the stronger of the hyperfine lines, $f$ is 2/3 and 2/3 of the 
total optical depth is due to this line, so these two factors cancel out.
This leads to a total \OHP\ column density of $2.4\times
10^{15}$~\cmsq. 

Similar to the analysis presented in \citet{menten+2010}, a
multi-component fit towards the spectra was done using the XCLASS tool
\citep[][and references therein]{comito+2005}, taking also the
blending of the two observed \OHP\ hyperfine lines into account.  The
background continuum temperature was estimated from the measured
continuum offset in the spectrum and the cosmic microwave background
temperature was used as the excitation temperature. For a reasonable
fit of the \OHP\ absorption, many more velocity components were
necessary than e.\ g.\ the ones reported by
\citet{garwood_dickey1989}. We therefore fit the same components
that \citet{menten+2010} used for \SHP\ and \THCHP, which also allows
a more consistent comparison of those hydrides with \OHP.  The result
of the fit is shown in Fig.~\ref{fig:spectrum} and the fit components
are given in Table \ref{t:ohpmodel}. 
Since the derived column densities are mostly determined by the
  observed line-to-continuum ratio, errors in the absolute calibration
  will not affect the results. Only uncertainties in the baseline level,
  as discussed in Sect.~\ref{sec:results}, might change the column densities
  by about 20\%.
The velocity components can
account for almost all of the \OHP\ absorption, missing only a few
velocity ranges, e.\ g.\ at $-64$~\kms\ and maybe at velocities beyond
$-116$~\kms, depending on the chosen baseline level. 
In order to compare \OHP\ column densities also with OH
and \OI\ results obtained with the ISO long wavelength spectrometer
\citep{vastel+2002, polehampton+2005}, we did additional \OHP\ fits
using those velocity parameters (see Table~\ref{t:ohpmodel}).
The total column densities for the different decompositions are between
  2.3 and $2.5 \times 10^{15}$~\cmsq.

The measured \OHP\ column densities $N$(\OHP) are shown in
Fig.~\ref{fig:cd-comparison} together with \SHP\ and \THCHP\ results
\citep{menten+2010}. $N$(\OHP) is of order $10^{14}$~\cmsq\ for a
large range of velocities and seems to follow more closely the \THCHP\
than the \SHP\ velocity structure. This is seen in more detail in
Fig.~\ref{fig:ratios} which shows comparisons of the column densities
of several molecules with \OHP. Loose correlations are found, the
strongest being with \CHP\ which is surprising since there is no
clear chemical link between the two. Interestingly, also for OH and CH
a close correlations was recently found in observations of translucent
clouds \citep{weselak+2010}. 
This correlation might be simply due to the fact that both hydrides
have a constant ratio to H$_2$.
 Note that for the comparison with \OI\ only velocities below
10~\kms\ were considered since \OI\ turns into emission at larger
velocities \citep{vastel+2002}.

\begin{table}
\caption{\OHP\ fit results using the velocity
components from the \SHP\ analysis and from \citet{garwood_dickey1989}.}
\label{t:ohpmodel} 
\begin{center}
 \begin{tabular}{rrr}
 \hline\hline 
 $V_{\mathrm{lsr}}$ & $\Delta V$ & $N$ \\ 
 (km~s$^{-1}$) & (km~s$^{-1}$) & ($10^{12}$~\cmsq) \\
\hline 
  \multicolumn{3}{l}{\citet{menten+2010} components:} \\
  64.0 & 19 & 257   \\ 
  38.5 &  5 &  6   \\ 
  32.0 &  5 & 24   \\ 
  22.0 &  5 & 44   \\ 
  17.0 &  5 & 29   \\ 
  12.0 &  6 & 22   \\ 
   8.2 &  5 & 88   \\ 
   2.5 &  5 & 103   \\ 
  -2.5 &  4 & 88   \\ 
  -8.7 &  5 & 88   \\ 
 -15.5 &  6 & 147   \\ 
 -20.0 &  5 & 96   \\ 
 -25.0 &  6 & 221   \\ 
 -29.5 &  4 & 59   \\ 
 -35.5 &  4 & 88   \\ 
 -41.3 &  3 & 74   \\ 
 -45.7 &  3 & 66   \\ 
 -49.0 &  3 & 66   \\ 
 -53.5 &  4 & 88   \\ 
 -58.7 &  4 & 96   \\ 
 -69.8 &  6 & 118   \\ 
 -76.0 &  4 & 81   \\ 
 -80.1 &  4 & 66   \\ 
 -84.3 &  5 & 96   \\ 
 -88.5 &  4 & 66   \\ 
 -93.0 &  4 & 74   \\ 
 -97.5 &  4 & 66   \\ 
-101.8 &  4 & 66   \\ 
-106.2 &  4 & 37   \\ 
-111.6 &  4 & 25   \\ 
-116.0 &  4 & 24   \\ 

 \hline
  \multicolumn{3}{l}{\citet{garwood_dickey1989} components:} \\
-108.0 &  7 &  51   \\ 
 -82.0 & 28 & 625   \\ 
 -52.0 & 17 & 257   \\ 
 -46.0 &  8 &  74   \\ 
 -24.4 & 14 & 515   \\ 
   1.1 & 19 & 441   \\ 
  15.7 &  7 &  15   \\ 
  31.0 & 21 &  74   \\ 
  52.8 & 11 &  44   \\ 
  66.7 & 16 & 221   \\ 

\hline
  \multicolumn{3}{l}{\citet{vastel+2002} components:} \\
-108.0 &  7 &  66   \\ 
 -92.0 & 14 & 294   \\ 
 -77.0 & 14 & 294   \\ 
 -61.5 &  7 & 147   \\ 
 -52.0 &  8 & 147   \\ 
 -44.0 &  8 & 147   \\ 
 -21.5 & 15 & 515   \\ 
  -3.5 & 12 & 294   \\ 
   5.5 & 12 & 221   \\ 
  31.0 & 21 &  59   \\ 
  52.8 & 11 &  44   \\ 
  66.7 & 16 & 221   \\ 

\hline
 \end{tabular}
 \end{center}
\end{table}

\section{Discussion and  conclusions}

\begin{table}
  \caption{Median column density ratios and ratio ranges of 
    different molecules with respect to \OHP\ 
    \citep{menten+2010,polehampton+2005,vastel+2002}.}
\label{t:ratios} 
\begin{center}
 \begin{tabular}{lrrr}
 \hline\hline 
  & \multicolumn{3}{c}{$N$(X)/$N$(\OHP)} \\
 \cline{2-4}
 \noalign{\smallskip}
 Molecule & Min. & Median & Max. \\ 
\hline 
 SH$^+$ &       0.03 &       0.13 &       1.28 \\
 CH$^+$ &       0.13 &       0.43 &       2.61 \\
     OH &      10.50 &      15.83 &     245.09 \\
     \OI\ &    1191.42 &    2587.08 &   11392.25 \\

\hline
 \end{tabular}
 \end{center}
\end{table}

Table~\ref{t:ratios} gives the column density ratios of several
molecules with \OHP.  For \SHP\ and \CHP\ the highest ratios are
reached towards the Sgr B2(M) local velocity of 64~\kms\ but for the
rest of the line-of-sight \OHP\ is more abundant. Ratios to OH and
atomic oxygen are found to be in the range of $10^{1-2}$ and
$10^{3-4}$, respectively. For velocities between --120 and --10~\kms,
\citet{neufeld+2000} give ortho-water column densities of about
$4\times 10^{15}$~\cmsq, hence the water to \OHP\ ratio in this
velocity range is about 40.

\citet{staeuber+2005} characterize \OHP\ as a pure X-ray tracer, more
likely to be enhanced by X-rays than by FUV radiation but their models
for higher density star formation environs cannot reproduce the low
OH/\OHP\ ratio observed here on the Sgr B2(M) line-of-sight. 
On the other hand,
the old \citet{glassgold_langer1976} and \citet{barsuhn_walmsley1977}
studies for low density, diffuse environs predict ratios in agreement
with the observed ones. \citet{vandishoeck_black1986} emphasize that
the formation rates of the simple oxygen-bearing molecules are
proportional to the cosmic-ray ionization rate. Their models of
diffuse clouds result in typical ratios between OH and \OHP\ of about
100, as typically found here, but the lines of sight that they 
model have significantly
lower column densities than the Sgr B2(M) intervening clouds.

Optical depths of individual velocity components of \OHP\ go up to 1.5
meaning that for the other THz lines of \OHP\ at 972 and 1033~GHz,
which are about a factor 4 and 2, respectively, stronger, optical
depths will become a problem in the analysis of the column densities.
Also the increasing number of hyperfine components will become
an issue in the interpretation of absorption of large velocity
ranges.  For extreme clouds, it might even be feasible to observe the
less abundant $^{18}$\OHP\ isotopologue.

The first detection of \OHP\ presented here, detected by its
absorption of a continuous velocity range on the Sgr B2(M)
line-of-sight, shows \OHP\ to be an abundant ingredient of diffuse
clouds, even more abundant than published chemical models suggest.
Although the 909~GHz line is close to atmospheric water absorption,
this detection shows that \OHP\ can be studied from the ground from
excellent sites such as the Chajnantor plateau in Chile, offering a
new observational tool to study basic oxygen chemistry in interstellar
clouds.

\acknowledgements{For this research, the XCLASS program
  (http://www.astro.uni-koeln.de/projects/schilke/XCLASS) was used,
  which accesses the CDMS (http://www.cdms.de) and JPL
  (http://spec.jpl.nasa.gov) molecular data bases. We thank Holger
  M\"uller for comments about the spectroscopy of \OHP\ and an
  anonymous referee for a thorough review of the paper.}

\bibliographystyle{aa}
\bibliography{apex-ohp}

\clearpage

\end{document}